\documentclass[12pt]{article}    
\usepackage{setspace}
\usepackage{hyperref}
\usepackage{graphics}
\usepackage{epsfig}
\usepackage{url}
\usepackage{amsmath}
\usepackage{mathrsfs}
\usepackage[utf8]{inputenc}
\usepackage{graphicx}
\usepackage{fancyhdr}
\usepackage[flushleft]{threeparttable}
\usepackage[nottoc]{tocbibind}
\usepackage[graphicx]{realboxes}
\usepackage{pdflscape}
\usepackage{rotating}
\usepackage{amssymb}
\usepackage{setspace}
\usepackage{hyperref}
\usepackage{graphics}
\usepackage{epsfig}
\usepackage{url}
\usepackage{amsmath}
\usepackage{mathrsfs}
\usepackage{amssymb}

\usepackage{amsmath}
\usepackage{bm}
\usepackage{amsfonts}
\usepackage{amsmath}
\usepackage{mathrsfs}
\usepackage{multirow}
\numberwithin{equation}{section}
\usepackage{tablefootnote}
\usepackage{booktabs}
\usepackage{multirow}
\usepackage{siunitx}
\usepackage[flushleft]{threeparttable}
\usepackage{color}
\usepackage{hyperref}
\usepackage{graphicx}
\usepackage{graphics}
\usepackage{adjustbox}
\usepackage{titlesec}

\usepackage{amssymb}

\def \R {{\rm I\kern -2.2pt R\hskip 1pt}}

\newcommand{\qed}{\rule{0.5em}{1.5ex}}

\usepackage{color}
\begin{document}
	
	\begin{center} { \large \sc  On Bivariate Pseudo-Logistic Distribution: Its Properties, Estimation and Applications}\\
		\vskip 0.1in {\bf Banoth Veeranna}\footnote{veerusukya40@gmail.com}
		\\
		\vskip 0.1in 
		School of Mathematics and Statistics, University of Hyderabad, Hyderabad, India. \\

		

	\end{center}
	
	\begin{abstract} 
		 The literature has covered the features and uses of the traditional univariate and bivariate logistic distributions in great detail. It is reasonable to wonder, though, if logistic marginals and conditionals could exhibit a similar behavior. A phenomenon that is comparable to both bivariate exponential and bivariate normal distributions. In this study, we will concentrate on bivariate distributions where one family of conditionals is marginal and the other family is of logistic type. Pseudo-logistic distributions are the name for such distributions. Research on conditionally specified models has revealed, however, that only in cases where the variables are independent will logistic marginals and both conditionals be of the logistic form occur. We talk about the features of distributional aspects and how they are built using the original. Both the original and the new conditioning regimes are used in two different ways. Possible generalizations are also considered. We also provide an example of a Pseudo-logistic model application.
	\end{abstract}
	
	\textbf{keywords:} Logistic distribution, Marginal and Conditional distribution, Pseudo-Poisson distribution, Maximum Likelihood Estimation, Likelihood ratio test.
	
	\bigskip

	\section{Introduction }  Two types of bivariate logistic distributions were introduced by Gumbel [1961], one of which had a (cumulative) distribution function is
	\begin{equation}
	F(x,y) = [1+e^{-x}+e^{-y}]^{-1}
	\end{equation}
	The regression curves and other key characteristics of this distribution were deduced by him. The absence of any parameter regarding the correlation between x and y significantly restricts the utility of this distribution.\\
	Following that, S.P. Satterthwaite and T.P. Hutchinson [1978] presented an extension of Gumbel's bivariate logistic distribution and performed an arbitrary power analysis on the generalized expression. A blend of bivariate extreme-value distributions could produce such a distribution.  A few fundamental characteristics of this distribution were also deduced. \\
	J. Filus and L. Filus [2006-2014] claim that a number of studies have examined what they refer to as "pseudo-exponential models," or models in which one marginal is exponential, and the other is the conditional distribution of the second variable assuming that the first variable's values are also exponential. also, we presented a bivariate pseudo-logistic distribution with one logistic form marginal density, let's say of X, and all conditional distributions of Y given X, also in logistic form. The scenario where the regression of Y given X is linear will receive special attention. whereby queries about parameter estimation for both whole and sub-models are addressed.\\
	In this section, we derived the linear form of the bivariate Pseudo-logistic distribution and its general properties. In section 2, we estimated the parameters using various methods, like the method of moments and maximum likelihood estimation. Also, we discuss the likelihood ratio test statistic for full and their sub-models in section 3. In section 4, we discussed the confidence intervals of the parameters of this model. Section 5 contains a little simulation study for the full and sub-model. We also apply the particular application of this model in section 6. Finally, section 7 contains some remarks and a conclusion part for this model.
	
\textbf{Definition:} A 2-dimensional random variable $\underline{X} = (X_1, X_2)$ or $(X, Y)$ is said to have a bivariate pseudo-logistic distribution, if there exists a location parameter $\mu$ and scale parameter $\sigma_0$ such that 
\begin{align*}
X \sim Logistic(\mu, \sigma_0)
\end{align*}
and a function of $\mu(x)$ such that
\begin{align*}
Y|X=x \sim Logistic(\mu(x), \sigma_1).
\end{align*}
Therefore the bivariate pseudo-logistic model with linear regression function, we assume that
\begin{align}
X \sim Logistic(\mu, \sigma_0)
\end{align}
and
\begin{align}
Y|X=x \sim Logistic(\mu(x), \sigma_1),
\end{align}
where $\mu(x) = \alpha + \beta x$, and the parameter space $\{(\mu, \sigma_0, \alpha, \beta, \sigma_1): \mu \geq 0, \sigma_0>0, \alpha>0, \beta \geq 0, \sigma_1>0 \}$.\\
The joint probability density function of X and Y is given by
\begin{align}
f_{X,Y}(x,y) = \frac{exp(-\frac{x-\mu}{\sigma_0})}{\sigma_0[1+exp(-\frac{x-\mu}{\sigma_0})]^2}\frac{exp(-\frac{y-\alpha-\beta x}{\sigma_1})}{\sigma_1[1+exp(-\frac{y-\alpha-\beta x}{\sigma_1})]^2} ; ~~~(x,y)\in \Re^2.
\end{align}
\subsection{Moments:}
The joint p.d.f. is given by
\begin{align*}
f_{X,Y}(x,y) = \frac{exp(-\frac{x-\mu}{\sigma_0})}{\sigma_0[1+exp(-\frac{x-\mu}{\sigma_0})]^2}\frac{exp(-\frac{y-\alpha-\beta x}{\sigma_1})}{\sigma_1[1+exp(-\frac{y-\alpha-\beta x}{\sigma_1})]^2} ; ~~~(x,y)\in \Re^2.
\end{align*}
Now,
\begin{align*}
E(X) = \mu,  ~~~~~~~ Var(X) = \frac{\sigma_0^2\pi^2}{3} 
\end{align*}
and
\begin{align*}
E(Y|X) = \alpha+\beta x, ~~~~~~~~ Var(Y|X) = \frac{\sigma_1^2\pi^2}{3}\\
E(Y) = E[E(Y|X)] = E[\alpha+\beta X] = \alpha + \beta\mu, \\
Var(Y) = E[Var(Y|X)] + Var[E(Y|X)] \\
= E[\frac{\sigma_1^2\pi^2}{3}]+Var[\alpha+\beta X]\\
Var(Y) = \frac{\pi^2}{3}(\sigma_1^2+\beta^2\sigma_0^2).
\end{align*}
Also,
\begin{align*}
E(XY) = E\{E(X.Y|X)\}= E[X.E(Y|X)] \\
= E[X(\alpha+\beta X)]=E[\alpha X+\beta X^2]\\
E(XY)= \alpha\mu+\beta(\mu^2+\frac{\sigma_0^2\pi^2}{3}).
\end{align*}
Therefore the covariance between X and Y is given by
\begin{align*}
Cov(X,Y)=E(XY) - E(X)E(Y) = \frac{\beta\pi^2\sigma_0^2}{3}.
\end{align*}
Also, the correlation coefficient is 
\begin{align*}
Corr(X,Y) = \rho = \frac{Cov(X,Y)}{\sqrt{Var(X)Var(Y)}}\\
\rho= \frac{\beta\sigma_0}{\sqrt{(\sigma_1^2+\beta^2\sigma_0^2)}}.
\end{align*}
\textbf{Case(1):-} When $\beta=0$, it follows that $\rho=0$, in fact, in this case X and Y are independent random variables.\\
\textbf{Case(2):-} If $\sigma_0 = \sigma_1$, then the correlation coefficient becomes
\begin{align*}
\rho=\frac{\beta}{\sqrt{(1+\beta^2)}}
\end{align*}
\textbf{Case(3):-} when $\sigma_0=1$, then the coefficient of correlation is
\begin{align*}
\rho=\frac{\beta}{\sqrt{(\sigma_1^2+\beta^2)}}
\end{align*}
\textbf{Case(4):-} If $\sigma_1=1$, then the correlation coefficient becomes
\begin{align*}
\rho= \frac{\beta\sigma_0}{\sqrt{(1+\beta^2\sigma_0^2)}}.
\end{align*}

\section{Statistical Inference}
\subsection{Method of Moments:}
The first moment is
\begin{align*}
E(X) = \mu\\
since~~ E(X) = \overline{X} \\
\Rightarrow  \hat{\mu} = M_1\\
since ~~M_1 = \frac{1}{n}\sum^n_{i=1}x_i, ~~~~~ M_2=\frac{1}{n}\sum^n_{i=1}y_i
\end{align*}
From simple linear model $\hat{\alpha} = \overline{Y} - \hat{\beta}\overline{X}$,
where $\hat{\beta} = \frac{S_{xy}}{S_{x}^2} = r_{xy}\frac{S_y}{S_x}$.
\begin{align*}
\hat{\alpha} = M_2-\frac{S_{12}}{S_1^2}M_1 , ~~~~ ~~~~ \hat{\beta} = \frac{S_{12}}{S_1^2}
\end{align*}
and, here $S_1^2 = Var(X)$ and $S_2^2 = Var(Y)$.
\begin{align*}
\hat{\sigma_0} = \frac{\sqrt{3}S_1^2}{\pi},~~~~~~ ~~~~ \hat{\sigma_1} = \frac{\sqrt{3(S_2^2-S_{12})}}{\pi}.
\end{align*}
\subsection{Maximum likelihood estimation:}
The given data of the form $(X_1,Y_1), ..., (X_n,Y_n)$ which are i.i.d. with common distribution of equation (3), then the likelihood function is as follows
\begin{align*}
L(\underline{\theta};\underline{x},\underline{y}) = \frac{exp(-\sum^n_{i=1}(\frac{x_i-\mu}{\sigma_0}))exp(-\sum_{i=1}^n(\frac{y_i-\alpha-\beta x_i}{\sigma_1}))}{\sigma_0^n\prod^n_{i=1}(1+exp(-\frac{x_i-\mu}{\sigma_0}))^2.\sigma_1^n\prod^n_{i=1}(1+exp(-\frac{y_i-\alpha-\beta x_i}{\sigma_1}))^2}
\end{align*}
where $\underline{\theta} = (\mu, \sigma_0, \alpha, \beta, \sigma_1)^T$. The corresponding log-likelihood function is given by
\begin{equation}
\begin{split}
l= log L(\underline{\theta}) &= \frac{n\mu}{\sigma_0}-\frac{n\overline{x}}{\sigma_0}-2\sum^n_{i=1}log(1+exp(-\frac{x_i-\mu}{\sigma_0}))-n log\sigma_0 \\
&+ \frac{n\alpha}{\sigma_1} + \frac{n\beta\overline{x}}{\sigma_1} - \frac{n\overline{y}}{\sigma_1} - 2\sum^n_{i=1}log(1+exp(-\frac{y_i-\alpha-\beta x_i}{\sigma_1})) -n log\sigma_1
\end{split}
\end{equation}
Now, we partial differentiating with respect to corresponding parameters and equating to zero,then we get following equations 
\begin{equation}
\frac{\partial l}{\partial \mu} = 0 \Rightarrow n-2\sum^n_{i=1}\frac{exp(-\frac{x_i-\mu}{\sigma_0})}{1+exp(-\frac{x_i-\mu}{\sigma_0})} =0,
\end{equation}
\begin{equation}
\frac{\partial l}{\partial \sigma_0} = 0 \Rightarrow \sigma_0 = \overline{x}-\mu-\frac{2}{n}\sum^n_{i=1}\frac{(x_i-\mu)exp(-\frac{x_i-\mu}{\sigma_0})}{1+exp(-\frac{x_i-\mu}{\sigma_0})},
\end{equation}
\begin{equation}
\frac{\partial l}{\partial \alpha} = 0 \Rightarrow n-2\sum^n_{i=1}\frac{exp(-\frac{y_i-\alpha-\beta x_i}{\sigma_1})}{1+exp(-\frac{y_i-\alpha-\beta x_i}{\sigma_1})} =0,
\end{equation}
\begin{equation}
\frac{\partial l}{\partial \beta} = 0 \Rightarrow n\overline{x}-2\sum^n_{i=1}\frac{x_i exp(-\frac{y_i-\alpha-\beta x_i}{\sigma_1})}{1+exp(-\frac{y_i-\alpha-\beta x_i}{\sigma_1})} =0,
\end{equation}
\begin{equation}
\frac{\partial l}{\partial \sigma} = 0 \Rightarrow \sigma_1 =  \overline{y}-\beta\overline{x}-\alpha-2\sum^n_{i=1}\frac{(y_i-\alpha-\beta x_i)exp(-\frac{y_i-\alpha-\beta x_i}{\sigma_1})}{1+exp(-\frac{y_i-\alpha-\beta x_i}{\sigma_1})}.
\end{equation}
The above equations must be solved numerically to obtain the estimated parameters, i.e., $\hat{\mu}, \hat{\sigma_0}, \hat{\alpha}, \hat{\beta}, \hat{\sigma_1}$.\\
\textbf{Note:-} We can use a fixed point iteration to find a better version of $\sigma_0$ for a given value of $\mu$, namely, $\sigma_{j+1} = h(\sigma_j)$. For any given value of $\sigma_0$, we can use Newton's method to find a better version of $\mu$, namely, $\mu_{j+1} = \mu_j-\frac{g(\mu_j)}{g'(\mu_j)}$, where $g'(\mu_j) = -2\sum^n_{i=1}\frac{exp(-\frac{x_i-\mu}{\sigma_0})}{1+exp(-\frac{x_i-\mu}{\sigma_0})}$.

\section{Likelihood Ratio Test}
As usual, the general form of a generalized likelihood ratio test statistic is of the form
\begin{align}
T = \frac{Sup_{\theta \in \Theta_0}L(\theta)}{Sup_{\theta \in \Theta}L(\theta)}.
\end{align}
Here, $\Theta_0$ is a subset of $\Theta$ and we envision testing $H_0: \theta \in \Theta_0$. We reject the null hypothesis for a small value of T. \\
In the following sub section we construct likelihood ratio tests for the simpler sub-models.
\subsection{Sub-model-1:}
For $\sigma_0 = \sigma_1$, equivalently, testing for $H_0: \sigma_0=\sigma_1$. The natural parameter space under the null hypothesis of the model is $\Theta_0=\{(\mu, \sigma_1, \alpha, \beta)^T: \mu \geq 0, \alpha>0, \beta \geq 0, \sigma_1>0 \}$. Besides the full model of the natural parameter space is $\Theta = \{(\mu, \sigma_0, \alpha, \beta, \sigma_1)^T: \mu \geq 0, \sigma_0>0, \alpha>0, \beta \geq 0, \sigma_1>0 \}$.\\
Under the $\Theta_0$, equation (4) will becomes
\begin{equation}
\begin{split}
l= log L(\underline{\theta}) &= \frac{n\mu}{\sigma_1}-\frac{n\overline{x}}{\sigma_1}-2\sum^n_{i=1}log(1+exp(-\frac{x_i-\mu}{\sigma_1}))-n log\sigma_1 \\
&+ \frac{n\alpha}{\sigma_1} + \frac{n\beta\overline{x}}{\sigma_1} - \frac{n\overline{y}}{\sigma_1} - 2\sum^n_{i=1}log(1+exp(-\frac{y_i-\alpha-\beta x_i}{\sigma_1})) -n log\sigma_1. 
\end{split}
\end{equation}
Now, we partial differentiating with respect to corresponding parameters and equating to zero,then we get following equations 
\begin{equation}
\frac{\partial l}{\partial \mu} = 0 \Rightarrow n-2\sum^n_{i=1}\frac{exp(-\frac{x_i-\mu}{\sigma_1})}{1+exp(-\frac{x_i-\mu}{\sigma_1})} =0,
\end{equation}
\begin{equation}
\frac{\partial l}{\partial \alpha} = 0 \Rightarrow n-2\sum^n_{i=1}\frac{exp(-\frac{y_i-\alpha-\beta x_i}{\sigma_1})}{1+exp(-\frac{y_i-\alpha-\beta x_i}{\sigma_1})} =0,
\end{equation}
\begin{equation}
\frac{\partial l}{\partial \beta} = 0 \Rightarrow n\overline{x}-2\sum^n_{i=1}\frac{x_i exp(-\frac{y_i-\alpha-\beta x_i}{\sigma_1})}{1+exp(-\frac{y_i-\alpha-\beta x_i}{\sigma_1})} =0,
\end{equation}
\begin{equation}
\begin{split}
\frac{\partial l}{\partial \sigma_1} = 0 \Rightarrow \sigma_1 &= \frac{\overline{y}+\overline{x}(1-\beta)-\alpha-\mu}{2}-\frac{1}{n}\sum^n_{i=1}\frac{(y_i-\alpha-\beta x_i)exp(-\frac{y_i-\alpha-\beta x_i}{\sigma_1})}{1+exp(-\frac{y_i-\alpha-\beta x_i}{\sigma_1})} \\ & -\frac{1}{n}\sum^n_{i=1}\frac{(x_i-\mu)exp(-\frac{x_i-\mu}{\sigma_1})}{1+exp(-\frac{x_i-\mu}{\sigma_1})}
\end{split}
\end{equation}
The above equations (3.3) - (3.6) are solved numerically, then we get M.L.E.'s of $\mu, \sigma_1, \alpha, \beta$ are called $\hat{\mu^*}, \hat{\sigma_1^*}, \hat{\alpha^*}, \hat{\beta^*}$, respectively. \\
Now, in the unrestricted parameter space $\Theta$. i.e., under the full model, the m.l.e.'s for $\mu, \sigma_0, \alpha, \beta, \sigma_1$ are obtained from equations (2.2) - (2.6). Let $\hat{\mu}, \hat{\sigma_0}, \hat{\alpha}, \hat{\beta}, \hat{\sigma_1}$ be the respective m.l.e.'s of $\theta$'s, then the generalized likelihood ratio test statistic defined in equation (10) will be 
\begin{equation}
T_1 = \frac{\frac{exp(-\sum^n_{i=1}\frac{x_i-\hat{\mu^*}}{\hat{\sigma_1^*}})exp(-\sum_{i=1}^n\frac{y_i-\hat{\alpha^*}-\hat{\beta^*} x_i}{\hat{\sigma_1^*}})}{(\hat{\sigma_1^*})^{2n}\prod^n_{i=1}[(1+exp(-\frac{x_i-\hat{\mu^*}}{\hat{\sigma_1^*}}))(1+exp(-\frac{y_i-\hat{\alpha^*}-\hat{\beta^*} x_i}{\hat{\sigma_1^*}}))]^2}}{\frac{exp(-\sum^n_{i=1}\frac{x_i-\hat{\mu}}{\hat{\sigma_0}})exp(-\sum_{i=1}^n\frac{y_i-\hat{\alpha}-\hat{\beta} x_i}{\hat{\sigma_1}})}{\hat{\sigma_0}^n\prod^n_{i=1}(1+exp(-\frac{x_i-\hat{\mu}}{\hat{\sigma_0}}))^2.\hat{\sigma_1}^n\prod^n_{i=1}(1+exp(-\frac{y_i-\hat{\alpha}-\hat{\beta} x_i}{\hat{\sigma_1}}))^2}}.
\end{equation}
\subsection{Sub-model-2:}
For $\sigma_0 = 1$, equivalently, testing for $H_0: \sigma_0=1$. The full model of the natural parameter space is $\Theta = \{(\mu, \sigma_0, \alpha, \beta, \sigma_1)^T: \mu \geq 0, \sigma_0>0, \alpha>0, \beta \geq 0, \sigma_1>0 \}$. Besides the natural parameter space under the null hypothesis of the model is $\Theta_0=\{(1,\mu, \sigma_1, \alpha, \beta)^T: \mu \geq 0, \alpha>0, \beta \geq 0, \sigma_1>0 \}$. \\
Under the $\Theta_0$, equation (4) will becomes
\begin{equation}
\begin{split}
l= log L(\underline{\theta}) &= n\mu-n\overline{x}-2\sum^n_{i=1}log(1+exp(-(x_i-\mu)) + \frac{n\alpha}{\sigma_1} \\
&+ \frac{n\beta\overline{x}}{\sigma_1} - \frac{n\overline{y}}{\sigma_1} - 2\sum^n_{i=1}log(1+exp(-\frac{y_i-\alpha-\beta x_i}{\sigma_1})) -n log\sigma_1. 
\end{split}
\end{equation}
Now, we partial differentiating with respect to corresponding parameters and equating to zero,then we get following equations 
\begin{equation}
\frac{\partial l}{\partial \mu} = 0 \Rightarrow n-2\sum^n_{i=1}\frac{exp(-(x_i-\mu))}{1+exp(-(x_i-\mu))} =0,
\end{equation}
\begin{equation}
\frac{\partial l}{\partial \alpha} = 0 \Rightarrow n-2\sum^n_{i=1}\frac{exp(-\frac{y_i-\alpha-\beta x_i}{\sigma_1})}{1+exp(-\frac{y_i-\alpha-\beta x_i}{\sigma_1})} =0,
\end{equation}
\begin{equation}
\frac{\partial l}{\partial \beta} = 0 \Rightarrow n\overline{x}-2\sum^n_{i=1}\frac{x_i exp(-\frac{y_i-\alpha-\beta x_i}{\sigma_1})}{1+exp(-\frac{y_i-\alpha-\beta x_i}{\sigma_1})} =0,
\end{equation}
\begin{equation}
\begin{split}
\frac{\partial l}{\partial \sigma_1} = 0 \Rightarrow \sigma_1 &= \overline{y}+\overline{x}\beta-\alpha-\frac{2}{n}\sum^n_{i=1}\frac{(y_i-\alpha-\beta x_i)exp(-\frac{y_i-\alpha-\beta x_i}{\sigma_1})}{1+exp(-\frac{y_i-\alpha-\beta x_i}{\sigma_1})}.
\end{split}
\end{equation}
The above equations (3.9) - (3.12) are solved numerically, then we get M.L.E.'s of $\mu, \sigma_1, \alpha, \beta$ are called $\hat{\mu^*}, \hat{\sigma_1^*}, \hat{\alpha^*}, \hat{\beta^*}$, respectively. \\
Now, in the unrestricted parameter space $\Theta$. i.e., under the full model, the m.l.e.'s for $\mu, \sigma_0, \alpha, \beta, \sigma_1$ are obtained from equations (2.2) - (2.6). Let $\hat{\mu}, \hat{\sigma_0}, \hat{\alpha}, \hat{\beta}, \hat{\sigma_1}$ be the respective m.l.e.'s of $\theta$'s, then the generalized likelihood ratio test statistic defined in equation (10) will be 
\begin{equation}
T_2 = \frac{\frac{exp(-\sum^n_{i=1}(x_i-\hat{\mu^*}))exp(-\sum_{i=1}^n\frac{y_i-\hat{\alpha^*}-\hat{\beta^*} x_i}{\hat{\sigma_1^*}})}{(\hat{\sigma_1^*})^{n}\prod^n_{i=1}[(1+exp(-(x_i-\hat{\mu^*})))(1+exp(-\frac{y_i-\hat{\alpha^*}-\hat{\beta^*} x_i}{\hat{\sigma_1^*}}))]^2}}{\frac{exp(-\sum^n_{i=1}\frac{x_i-\hat{\mu}}{\hat{\sigma_0}})exp(-\sum_{i=1}^n\frac{y_i-\hat{\alpha}-\hat{\beta} x_i}{\hat{\sigma_1}})}{\hat{\sigma_0}^n\prod^n_{i=1}(1+exp(-\frac{x_i-\hat{\mu}}{\hat{\sigma_0}}))^2.\hat{\sigma_1}^n\prod^n_{i=1}(1+exp(-\frac{y_i-\hat{\alpha}-\hat{\beta} x_i}{\hat{\sigma_1}}))^2}}.
\end{equation}
\subsection{Sub-model-3:}
For $\sigma_1 = 1$, equivalently, testing for $H_0: \sigma_1=1$. The natural parameter space under the null hypothesis of the model is $\Theta_0=\{(\mu, \sigma_0, \alpha, \beta)^T: \mu \geq 0, \alpha>0, \beta \geq 0, \sigma_0>0 \}$. Besides the full model of the natural parameter space is $\Theta = \{(\mu, \sigma_0, \alpha, \beta, \sigma_1)^T: \mu \geq 0, \sigma_0>0, \alpha>0, \beta \geq 0, \sigma_1>0 \}$.\\
Under the $\Theta_0$, equation (4) will becomes
\begin{equation}
\begin{split}
l= log L(\underline{\theta}) &= \frac{n\mu}{\sigma_0}-\frac{n\overline{x}}{\sigma_0}-2\sum^n_{i=1}log[1+exp(-\frac{x_i-\mu}{\sigma_0})]-n log\sigma_0 \\
&+ n\alpha + n\beta\overline{x}-n\overline{y} - 2\sum^n_{i=1}log[1+exp(-(y_i-\alpha-\beta x_i))]. 
\end{split}
\end{equation}
Now, we partial differentiating with respect to corresponding parameters and equating to zero,then we get following equations 
\begin{equation}
\frac{\partial l}{\partial \mu} = 0 \Rightarrow n-2\sum^n_{i=1}\frac{exp(-\frac{x_i-\mu}{\sigma_0})}{1+exp(-\frac{x_i-\mu}{\sigma_0})} =0,
\end{equation}
\begin{equation}
\begin{split}
\frac{\partial l}{\partial \sigma_0} = 0 \Rightarrow \sigma_0 &= \overline{x}-\mu-\frac{2}{n}\sum^n_{i=1}\frac{(x_i-\mu)exp(-\frac{x_i-\mu}{\sigma_0})}{1+exp(-\frac{x_i-\mu}{\sigma_0})}
\end{split}
\end{equation}
\begin{equation}
\frac{\partial l}{\partial \alpha} = 0 \Rightarrow n-2\sum^n_{i=1}\frac{exp(-(y_i-\alpha-\beta x_i))}{1+exp(-(y_i-\alpha-\beta x_i))} =0,
\end{equation}
\begin{equation}
\frac{\partial l}{\partial \beta} = 0 \Rightarrow n\overline{x}-2\sum^n_{i=1}\frac{x_i exp(-(y_i-\alpha-\beta x_i))}{1+exp(-(y_i-\alpha-\beta x_i))} =0,
\end{equation}

The above equations (3.15) - (3.18) are solved numerically, then we get M.L.E.'s of $\mu, \sigma_0, \alpha, \beta$ are called $\hat{\mu^*}, \hat{\sigma_0^*}, \hat{\alpha^*}, \hat{\beta^*}$, respectively. \\
Now, in the unrestricted parameter space $\Theta$. i.e., under the full model, the m.l.e.'s for $\mu, \sigma_0, \alpha, \beta, \sigma_1$ are obtained from equations (2.2) - (2.6). Let $\hat{\mu}, \hat{\sigma_0}, \hat{\alpha}, \hat{\beta}, \hat{\sigma_1}$ be the respective m.l.e.'s of $\theta$'s, then the generalized likelihood ratio test statistic defined in equation (10) will be 
\begin{equation}
T_3 = \frac{\frac{exp(-\sum^n_{i=1}\frac{x_i-\hat{\mu^*}}{\hat{\sigma_0^*}})exp(-\sum_{i=1}^n(y_i-\hat{\alpha^*}-\hat{\beta^*} x_i))}{(\hat{\sigma_0^*})^{n}\prod^n_{i=1}[(1+exp(-\frac{x_i-\hat{\mu^*}}{\hat{\sigma_0^*}}))(1+exp(-(y_i-\hat{\alpha^*}-\hat{\beta^*} x_i)))]^2}}{\frac{exp(-\sum^n_{i=1}\frac{x_i-\hat{\mu}}{\hat{\sigma_0}})exp(-\sum_{i=1}^n\frac{y_i-\hat{\alpha}-\hat{\beta} x_i}{\hat{\sigma_1}})}{\hat{\sigma_0}^n\prod^n_{i=1}(1+exp(-\frac{x_i-\hat{\mu}}{\hat{\sigma_0}}))^2.\hat{\sigma_1}^n\prod^n_{i=1}(1+exp(-\frac{y_i-\hat{\alpha}-\hat{\beta} x_i}{\hat{\sigma_1}}))^2}}.
\end{equation}

\section{The parameter's confidence intervals}
We are aware that constructing a confidence interval for $\mu$ and $\sigma$ with a given confidence coefficient of $100(1-\alpha)\%$ is challenging, even for univariate logistic distribution. Lower and upper bounds for the confidence interval with a given confidence coefficient can be established by interpolating in tables of the central $\chi^2$-distribution using the relationship between the logistic and $\chi^2$-distribution. We consult section 4.7.3 of Johnson, Kemp, and Kortz's work for additional details on creating a confidence interval for the logistic distribution.
The Wald method of creating confidence intervals will be the topic of this note. In general, the Wald confidence interval is given for every parameter $\theta$, and the related point estimator $\hat{\theta}$(say) is given by
\begin{align}
\hat{\theta} \pm Z_{\alpha/2}S.E(\hat{\theta}),
\end{align}
where S.E.($\hat{\theta}$) is the standard error of the estimator $\hat{\theta}$, and $Z_{\alpha/2}$ is the $100(1-\alpha)\%$ of the standard normal distribution. Furthermore, take note of the weak coverage qualities of the Wald confidence interval for small sample sizes.\\
Given that the pseudo-logistic marginal distribution of X has logistic parameters, $\mu$ and $\sigma_0$, a confidence interval for $\mu$ and $\sigma_0$ can be obtained using the current method (which relies on the link between the logistic and $\chi^2$-distribution). Confidence intervals for $\alpha$, $\beta$, and $\sigma_1$ can be found using the previously mentioned Wald approach. The behavior of the Wald confidence interval for the parameters for the small and big sample sizes will be examined in the ensuing subsections.

\section{Simulation study}
			Because of the marginal and conditional structure of the model, simulating from pseudo models is simple. Here, we present a straightforward simulation approach using linear regression for the bivariate pseudo-logistic model. We have simulated 10,000 data sets of sample size $n= 30, 50, 100, 200, 500$ from the following full and sub-model:
			Tables 1 and 2 provide the related moment estimations and M.L.Es, along with their bootstrapped standard errors and confidence intervals (CI) for the entire and sub-models, respectively.
			We provide a general summary of the Tables with the following observations. It is observed that as sample size increases, the Pearson Correlation (PC) converges to the population correlation, and the moment and m.l.e.'s standard error (SE) drop. Additionally, when compared to the confidence interval created using moment estimators, the Wald confidence interval generated using MLE estimators has the lowest length. interval constructed using moment estimators.

		\begin{figure}[htp]
			\centering
			\includegraphics[width=17cm]{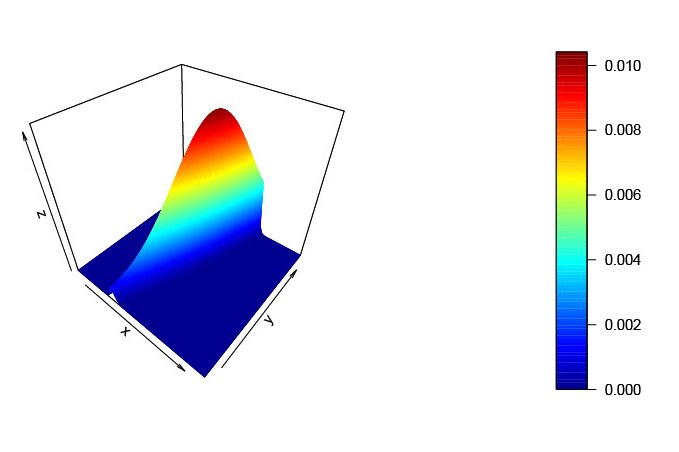}
			\caption{Bivariate density plot for pseudo-logistic model of the parameters values  $\mu=2, \sigma_0=3, \alpha=1, \beta=3$ and $\sigma_1=2$.}
			\label{fig17}
		\end{figure}

			\begin{table}
				\caption{{Simulation for Full Model}} 
				\label{Table_sub_I}
				\normalsize
				\centering 
				\resizebox{\textwidth}{!}{
					\begin{tabular}{ |c|c|c |c| c| c| c| c| c| c| c| c|} 
						\hline
						Sample Size & Parameters & MLE & SE(MLE) & Bias(MLE) & MM & SE(MM) & Bias(MM) & CI(MLE) & CI(MM) & PC & -2LogL \\
						\hline
						\multirow{5}{1em}{30} & $\mu$ & 2.056 & 1.006 & 0.056 & 1.656 & 0.0305 & 0.344 & (0.009, 5.915) & (0.143, 3.883) & \multirow{5}{3em}{0.9748} & \multirow{5}{5em}{358.7302}\\
						& $\sigma_0$ & 3.186 & 0.491 & 0.186 & 3.223 & 0.0147& 0.223 & (2.078, 5.900) & (2.105, 3.917) & & \\ 
						& $\alpha$ & 1.893 & 0.713 & 0.893 & 1.705 & 0.0228 & 0.705 & (-0.437, 8.429) & (-0.234, 2.648) & & \\ 
						& $\beta$ & 2.722 & 0.116 & 0.278 & 2.636 & 0.0040 &  0.364 & (2.196, 92.706) & (2.645, 3.142) & & \\
						& $\sigma_1$ & 2.221 & 0.346 & 0.221 & 7.439 & 0.0372 & 5.439 &(1.374, 288.808) & (5.413, 10.092) & & \\
						\hline
						\multirow{5}{1em}{50} & $\mu$ & 2.628& 0.645 & 0.628 & 2.753 & 0.0244 & 0.753 & (0.556, 3.564) & (0.527, 3.542) & \multirow{5}{3em}{0.975} & \multirow{5}{5em}{549.3604}\\
						& $\sigma_0$ & 2.622 & 0.310 & 0.378 & 2.572 & 0.0118& 0.428 & (2.310, 3.735) & (2.284, 3.760) & & \\ 
						& $\alpha$ & 1.317 & 0.477 & 0.317 & 1.610 & 0.0173 & 0.610 & (-0.106, 2.071) & (-0.035, 2.194) & & \\ 
						& $\beta$ & 2.911 & 0.094& 0.089 & 2.829 & 0.0033 &  0.171 & (2.761, 3.224) & (2.732, 3.131) & & \\
						& $\sigma_1$ & 1.729 & 0.204 & 0.271 & 6.264 & 0.0300 & 4.264 &(1.512, 21.723) & (5.816, 9.500) & & \\
						\hline
						\multirow{5}{1em}{100} & $\mu$ & 3.003 & 0.465 & 1.003& 2.973 & 0.0175 & 0.973 & (0.938, 2.961) & (0.879, 3.011) & \multirow{5}{3em}{0.9758} & \multirow{5}{5em}{1164.826}\\
						& $\sigma_0$ & 2.671 & 0.222 & 0.329 & 2.632 & 0.0082& 0.368 & (2.537, 3.473) & (2.512, 3.511) & & \\ 
						& $\alpha$ & 0.716 & 0.485 & 0.284 & 0.552 & 0.0127 & 0.448 & (0.268, 1.787) & (0.297, 1.874) & & \\ 
						& $\beta$ & 3.050 & 0.085 & 0.050 & 3.041 & 0.0022 &  0.041 & (2.875, 3.133) & (2.839, 3.105) & & \\
						& $\sigma_1$ & 2.293 & 0.194 & 0.293 & 7.088 & 0.0204 & 5.088 &(1.675, 2.338) & (6.394, 8.907) & & \\
						\hline
						\multirow{5}{1em}{200} & $\mu$ & 1.892 & 0.371 & 0.108 & 1.656 & 0.0119 & 0.125 & (1.274, 2.689) & (1.277, 2.717) & \multirow{5}{3em}{0.9762} & \multirow{5}{5em}{2271.648}\\
						& $\sigma_0$ & 3.017 & 0.178 & 0.017 & 2.971 & 0.0057& 0.029 & (2.671, 3.349) & (2.660, 3.382) & & \\ 
						& $\alpha$ & 0.879 & 0.225 & 0.121 & 825 & 0.0087 & 0.175 & (0.483, 1.521) & (0.503, 1.570) & & \\ 
						& $\beta$ & 2.962 & 0.041 & 0.038 & 2.922 & 0.0015 &  0.078 & (2.913, 3.089) & (2.895, 3.078) & & \\
						& $\sigma_1$ & 1.759 & 0.105 & 0.241 & 7.330 & 0.0143 & 5.330 &(1.758, 2.215) & (6.750, 8.579) & & \\
						\hline
						\multirow{5}{1em}{500} & $\mu$ & 1.982 & 0.228 & 0.018 & 2.054 & 0.0076 & 0.059 & (1.551, 2.441) & (1.525, 2.465) & \multirow{5}{3em}{0.976} & \multirow{5}{5em}{5772}\\
						& $\sigma_0$ & 2.937 & 0.109 & 0.063 & 2.902 & 0.0039& 0.098 & (2.764, 3.221) & (2.748, 3.227) & & \\ 
						& $\alpha$ & 1.014 & 0.168 & 0.014 & 1.096 & 0.0055 & 0.096 & (0.663, 1.315) & (0.678, 1.343) & & \\ 
						& $\beta$ & 2.949 & 0.030 & 0.051 & 2.930 & 0.0010 &  0.070 & (2.945, 3.056) & (2.934, 3.053) & & \\
						& $\sigma_1$ & 2.019 & 0.075 & 0.019 & 7.205 & 0.00962 & 5.205 &(1.870, 2.151) & (7.012, 8.192) & & \\
						\hline
					\end{tabular}
				}
				
			\end{table}

		\begin{table}[ht]
			\centering
			\caption{Simulation for Sub-Model I}
			\normalsize
			\centering 
			\resizebox{\textwidth}{!}{
			\begin{tabular}{ |c|c|c |c| c| c| c| c| } 
				\hline
				Sample Size & Parameters & MLE & SE(MLE) & Bias(MLE)  & CI(MLE) & PC & -2LogL \\
				\hline
				\multirow{4}{1em}{30} & $\mu$ & 2.046 & 1.002 & 0.046 & (0.118, 8.828) & \multirow{4}{3em}{0.9456} & \multirow{4}{5em}{383.098}\\
				& $\sigma_0$ & 3.257& 0.357 & 0.257 & (2.329, 401.286) &  & \\ 
				& $\alpha$ & 2.352 & 1.049 & 1.352 & (-1.190, 3.387) &  & \\ 
				& $\beta$ & 2.582 & 0.171 & 0.418 & (2.595, 9.558) &  & \\
				\hline
				\multirow{4}{1em}{50} & $\mu$ & 2.626 & 0.643 & 0.626 & (0.589, 3.546) & \multirow{4}{3em}{0.9464} & \multirow{4}{5em}{589.9109}\\
				& $\sigma_0$ & 2.608 & 0.217 & 0.392 & (2.482, 3.552) &  & \\ 
				& $\alpha$ & 1.477 & 0.719 & 0.477 & (-0.543, 2.597) &  & \\ 
				& $\beta$ & 2.866 & 0.142 & 0.134 & (2.680, 3.303) &  & \\
				\hline
				\multirow{4}{1em}{100} & $\mu$ & 2.989 & 0.515 & 0.989 & (0.941, 3.175) & \multirow{4}{3em}{0.9479} & \multirow{4}{5em}{1250.464}\\
				& $\sigma_0$ & 3.054& 0.181 & 0.054 & (2.657, 3.392) &  & \\ 
				& $\alpha$ & 0.589 & 0.669 & 0.411 & (-0.102, 2.184) &  & \\ 
				& $\beta$ & 3.074 & 0.117 & 0.074 & (2.798, 3.210) &  & \\
				\hline
				\multirow{4}{1em}{200} & $\mu$ & 1.891 & 0.354 & 0.109 & (1.276, 2.699) & \multirow{4}{3em}{0.9486} & \multirow{4}{5em}{2436.375}\\
				& $\sigma_0$ & 2.832 & 0.119 & 0.168 & (2.750, 3.234) &  & \\ 
				& $\alpha$ & 0.813 & 0.356 & 0.187 & (0.225, 1.783) &  & \\ 
				& $\beta$ & 2.942 & 0.064 & 0.058 & (2.870, 3.133) &  & \\
				\hline
				\multirow{4}{1em}{500} & $\mu$ & 1.983 & 0.231 & 0.017 & (0.553, 2.442) & \multirow{4}{3em}{0.9482} & \multirow{4}{5em}{6177.807}\\
				& $\sigma_0$ & 2.983 & 0.079 & 0.017 & (2.836, 3.155) &  & \\ 
				& $\alpha$ & 1.018 & 0.249 & 0.018 & (0.495, 1.472) &  & \\ 
				& $\beta$ & 2.925 & 0.045 & 0.075 & (2.918, 3.084) &  & \\
				\hline
			\end{tabular}
		}
		\end{table}	
			
\newpage
\section{Applications:}
We consider a data set in which the source of the data from the scores obtained by $n=87$ college students on the College Level Examination Program (CLEP) subtests $X_1$ and the College Qualification Test (CQT) subtests $X_2$ and $X_3$ are mentioned in Johnson and Wichern (2007, p.228) is data set given in Table 5.2.
The numerical method was used to derive the maximum likelihood estimates of the parameters for the bivariate pseudo-logistic model. Estimates, standard errors, AIC, BIC, and Pearson Correlation (PC) were calculated for each full model and sub-model and are shown in Tables 3 and 4, respectively.

\begin{table}[ht]
	\centering
	\caption{Application for Full Model}
	\normalsize
	\centering 
	\resizebox{\textwidth}{!}{
	\begin{tabular}{ |c|c|c |c| c| c| c| c| c| } 
		\hline
		Sample Size & MLE & SE(MLE) & MM  & -2LogL & PC & AIC & BIC \\
		\hline
		\multirow{5}{1em}{87}  & 52.517 & 0.825 & 52.659 &  \multirow{5}{4em}{1086.441} & \multirow{5}{3em}{0.606} & \multirow{5}{5em}{1096.441} & \multirow{5}{5em}{1108.771} \\
		& 4.387 & 0.388 & 4.202 & & & &  \\
		& 4.775 & 0.817 & 5.228 & & & &  \\
		& 0.387 & 0.053 & 0.378 & & & &  \\
		& 2.241 & 0.197 & 0.595 & & & &  \\
		\hline
		
	\end{tabular}
}
\end{table}
\begin{table}[ht]
	\centering
	\caption{Application for Sub-Model I}
	\begin{tabular}{ |c|c|c |c| c| c| c| c|  } 
		\hline
		Sample Size & MLE & SE(MLE)  & -2LogL & PC & AIC & BIC \\
		\hline
		\multirow{4}{1em}{87}  & 52.459 & 0.682  &  \multirow{4}{4em}{1114.631} & \multirow{4}{3em}{0.606} & \multirow{4}{5em}{1122.631} & \multirow{4}{5em}{1140.495} \\
		& 3.337 & 0.212  & & & &  \\
		& 4.933 & 0.904  & & & &  \\
		& 0.384 & 0.074  & & & &  \\
		
		\hline
		
	\end{tabular}
	
\end{table}
		
\section{Conclusion}		
We have developed flexible models known as bivariate pseudo-logistic distributions by taking into consideration bivariate models in which one marginal distribution is assumed to be of the logistic form while the conditional distributions of the second variable, given the first, are also assumed to be of the logistic form. For these models, investigation has been done into distributional and inferential problems. The different bivariate logistic models that have been introduced in the literature may be replaced by the models that are covered in this study. The pseudo models' clear form makes it possible to fit the model and estimate parameters with ease, as well as to simulate it simply. However, we would contend that they—as well as their extensions to higher dimensions and permuted variations of them—are not a cure-all.

\section{Acknowledgement(s)}
I thank the Ministry of Tribal Affairs-National Fellowship Scheme For Higher Education of ST Students(NFST) for providing me with a Junior Research Fellowship (JRF) and Senior Research Fellowship (SRF) (award no: 201819-NFST-TEL-00347).

		\end{document}